\documentclass{aa}  
\usepackage[colorlinks=true,urlcolor=blue,citecolor=blue,linkcolor=blue]{hyperref}
\usepackage{graphicx}
\usepackage{subfig}
\usepackage{txfonts}
\begin{document}

\title{SunnyNet: A neural network approach to 3D non-LTE radiative transfer}

\author{Bruce A. Chappell
	\inst{1,2}
	\and Tiago M.D. Pereira
	\inst{1,2}
	}
\institute{Rosseland Centre for Solar Physics, University of Oslo, P.O. Box 1029 Blindern, NO--0315 Oslo, Norway
\and
Institute of Theoretical Astrophysics, University of Oslo, P.O. Box 1029 Blindern, NO--0315 Oslo, Norway}
\date{\today}

   \date{}

  \abstract
   {Computing spectra from 3D simulations of stellar atmospheres when allowing for departures from local thermodynamic equilibrium (non-LTE) is computationally very intensive.}
   {We develop a machine learning based method to speed up 3D non-LTE radiative transfer calculations in optically thick stellar atmospheres.}
   {Making use of a variety of 3D simulations of the solar atmosphere, we trained a convolutional neural network, SunnyNet, to learn the translation from LTE to non-LTE atomic populations. Non-LTE populations computed with an existing 3D code were considered as the true values. The network was then used to predict non-LTE populations for other 3D simulations, and synthetic spectra were computed from its predicted non-LTE populations. We used a six-level model atom of hydrogen and H$\alpha$ spectra as test cases.}
   {SunnyNet gives reasonable predictions for non-LTE populations with a dramatic speedup of about 10$^5$ times when running on a single GPU and compared to existing codes. When using different snapshots of the same simulation for training and testing, SunnyNet's predictions are within 20-40\% of the true values for most points, which results in average differences of a few percent in H$\alpha$ spectra. Predicted H$\alpha$ intensity maps agree very well with existing codes. Most importantly, they show the telltale signs of 3D radiative transfer in the morphology of chromospheric fibrils. The results are not as reliable when the training and testing are done with different families of simulations. SunnyNet is open source and publicly available.}
  {}

   \keywords{Radiative transfer -- Methods: numerical -- Line: formation -- Sun: atmosphere -- Stars: atmospheres}

   \maketitle

\section{Introduction}

Forward modeling of stellar spectra using 3D models of atmospheres has become an invaluable tool. Its applications range from ushering in a new era of solar and stellar abundance determinations \citep[see][and references therein]{Asplund2005}, understanding the dynamics of the solar atmosphere \citep[e.g.,][]{Martinez-Sykora:2009, Danilovic:2010, Pereira:2015, Leenaarts:2015, cbh24, Schmit:2017, Tei:2020}, and guiding the interpretation of observations, in particular from new instruments or observatories \citep[e.g.,][]{Pereira:2013aa, QuinteroNoda:2017, Bjorgen:2018, TrujilloBueno:2018, Riethmuller:2019, Jafarzadeh:2019}. 

For cool stars, the atmospheres are optically thick at most of the emitted wavelengths and, in particular for the outer atmospheric layers, the plasma is decoupled from the radiation field and therefore the handy approximation of local thermodynamic equilibrium (LTE) no longer holds. This means that to obtain accurate synthetic spectra, one must solve the full 3D radiative transfer problem in non-LTE (NLTE). Such calculations require vast amounts of atomic data: in addition to the usual ingredients of spectral synthesis, one must also model in detail many atomic levels and transitions (not only the ones involved in the line(s) of interest) and include the cross sections that govern collisional excitation and de-excitation from atomic levels at different temperatures. The NLTE problem is nonlocal, as radiation traveling large distances can couple to different parts of the atmosphere. Therefore, it becomes computationally very expensive. For studies of the solar chromosphere in particular, where high spatial resolution is important to resolve small-scale phenomena, the cost of 3D NLTE spectral synthesis becomes prohibitively expensive, requiring supercomputers for even simple applications. 

Presently, there are several 3D NLTE codes used for cool-star spectral synthesis, for example Multi3D \citep{Leenaarts:2009}, PHOENIX/3D \citep{Hauschildt:2010}, and PORTA \citep{Stepan:2013}, to name a few that are also massively parallel. These codes solve the radiative transfer equation iteratively to obtain estimates for the atomic level populations and radiation field that are consistent in all locations in the 3D atmosphere. Typically, variations of a procedure called $\Lambda$ iteration are used. The result is an estimate of the source function $S_\lambda$, the ratio between the extinction coefficient and the emissivity at a given wavelength. Once $S_\lambda$ is known, it is straightforward to solve the radiative transfer equation for a given ray (represented by a direction in optical depth $\tau_\lambda$):
\begin{equation}
    \label{eq:rte}
    \frac{\mathrm{d} I_\lambda}{\mathrm{d} \tau_\lambda} = I_\lambda - S_\lambda,
\end{equation}
which is typically done by integrating the equation over the depth points along the ray (the so-called formal solution). The bulk of the computational effort is therefore spent in the iterative procedure to obtain the estimate of the source function, which also involves obtaining the atomic level populations and the angle-averaged radiation field.

In this work, we present a novel approach to accelerate 3D NLTE radiative transfer calculations in atmospheres of cool stars, focusing on the solar chromosphere as a main application. Instead of following the traditional approach through a $\Lambda$ iteration to obtain the source function, we instead used a neural network to learn the mapping between LTE and NLTE populations of an atomic species of interest. We run the Multi3D code on a training set of atmospheres for a given model atom, and treat the results as the absolute true mapping between LTE and NLTE populations, which is used to predict the NLTE populations for an arbitrary atmosphere. We then use the estimated NLTE populations to compute the synthetic spectra. The idea of using neural networks to learn the mapping between atmosphere and NLTE spectra was already used by \citet{osborne} to invert observations of flares. And this approach also relates to a growing body of work \citep[e.g.,][]{AsensioRamos:2019, Beck:2019, Milic:2020, Gafeira:2021} dedicated to speeding up spectropolarimetric inversions from observations, a procedure that also involves repeatedly solving the radiative transfer equation and is computationally very demanding.

What most distinguishes our approach from previous work is that our network considers the full 3D problem and is not limited to the 1.5D approximation of treating each column from a 3D box as an independent plane-parallel atmosphere (neglecting inclined rays). As shown by \citet{Leenaarts:2012} for the H$\alpha$ line, full 3D NLTE is needed to model several lines formed in the solar chromosphere, in particular the radiation near the line cores.

\section{SunnyNet}

\subsection{Motivation and overall design}

In this section we detail the design of our neural network approach to speed up 3D NLTE calculations, which we named SunnyNet. The overarching goal is a fast method to compute synthetic spectra from a 3D model atmosphere, for a given atomic transition and viewing angle. Existing methods can already compute accurate solutions to this problem, but are notoriously slow. Hence our focus here is on speed of computation. 

Since the end goal is the calculation of spectra, one can envisage a procedure to learn the mapping from atmospheric properties (e.g., temperature, density, electron density, velocities) to the intensity spectra directly. Indeed this is the approach taken by \citet{osborne}. However, using this approach one needs to train the network in advance for each spectral line of interest, and for each viewing angle, since intensity is direction-dependent. Adding more spectral lines or different viewing angles would require re-running the training procedure. This can be time-consuming, since the training of neural networks is the most intensive part of the process. To allow for more flexibility, we instead designed SunnyNet to learn the mapping between LTE and NLTE atomic populations. 

The atomic populations are a more fundamental quantity that describe the state of matter for a given atom. They are independent of viewing angle and also allow the synthesis of any transition between the levels of the chosen model atom. In this approach we use LTE populations as input to SunnyNet, instead of the atmospheric properties themselves. This is feasible because in LTE the populations, through the Saha-Boltzmann distribution, are a function of the atmospheric properties alone and already reflect the state of the atmosphere. Having the input and output of SunnyNet to have the same dimensions (i.e., populations as a function of atomic level and spatial position) also allows for a simpler algorithm. The final predicted spectra can be computed by doing a single formal solution using the output NLTE populations.

Throughout this work, we make use of 3D model atmospheres ran with from the Bifrost code \citep{Gudiksen:2011}. We ran Multi3D in some of the atmospheres to obtain what we adopt as the true mapping between LTE and NLTE populations. We use a 5-level plus continuum model atom of hydrogen (6 levels in total), although the procedure can be generalized to any other model atoms.

Our implementation of SunnyNet was developed primarily in Python using the PyTorch library \citep{PyTorch}. The code is freely available\footnote{\url{https://github.com/bruce-chappell/SunnyNet}} and licensed under a BSD license. The code version used in this work is given by \citet{SunnyNet}.

\subsection{Convolutional neural networks}
Convolutional neural networks (CNNs) have become increasingly popular tools for learning complex mappings between a set of inputs and outputs. They learn a mapping $h$:

\begin{equation}
    h\left( \boldsymbol{X_i}, \boldsymbol{\hat{\theta}} \right) = \boldsymbol{\tilde{y}_i}
    ,
\end{equation}
which approximates the true function $f$:
\begin{equation}
    f\left( \boldsymbol{X_i} \right) = \boldsymbol{y_i}
    ,
\end{equation}
by minimizing the total loss function $L$:
\begin{equation}
    \min\limits_{\boldsymbol{\hat{\theta}}} L\left( \boldsymbol{\hat{\theta}} \right) = \min\limits_{\boldsymbol{\hat{\theta}}} \sum_{i=1}^{n} l_{i}\left( \boldsymbol{\hat{\theta}}; \boldsymbol{y_i}, \boldsymbol{\tilde{y}_i} \right)
    .
\end{equation}

This is done through an iterative process where the input data is passed forward through the model, a prediction is made, and the loss is calculated. We then implement the backpropagation algorithm of \citet{backprop}. This algorithm leverages the chain rule and calculates the partial derivative of the loss with respect to each weight and bias, thus telling us how much each parameter contributes to the loss. We then take a step in the parameter space in the direction of the negative gradient of the loss using the gradient descent algorithm of \citet{murphy} and update the parameters. By repeating this process until the loss is minimized, we fine-tuned the weights and biases, $\boldsymbol{\hat{\theta}}$, of our model $h$ and arrive at a best possible approximation for $f$.

While most image processing networks utilize 2D convolutions, SunnyNet utilizes 3D and 1D convolutions. In a 3D convolution layer, the weights are sets of filters with shape $(c, z, x, y)$ where $c$ is the number of channels and the remaining quantities are spatial dimensions. These filters convolve in three dimensions across the input transforming the 4D input into a 1D output. A 1D convolution layer acts similarly, in that it has filters of shape $(c, z)$ that convolve across one dimension.

\begin{figure}
    \centering
    \includegraphics[scale = 0.25]{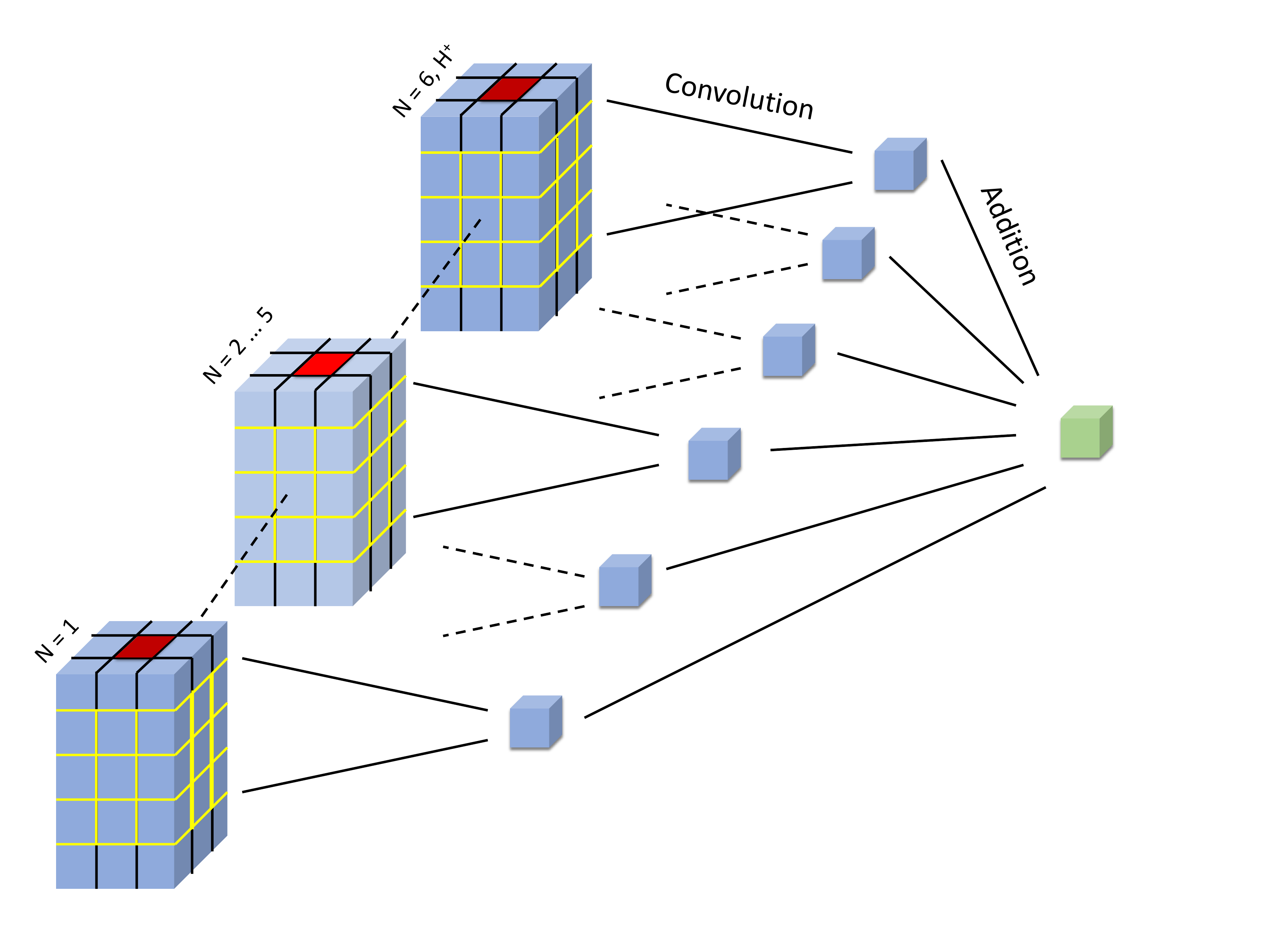}
    \caption{Visualization of 3D convolution for $3\times3$ input data.}
    \label{3d conv}
\end{figure}

\subsection{Data structure and $z$ scale}

To build our datasets we made use of atomic populations computed with Multi3D. With our six-level model hydrogen atom, the shape of our input data is therefore $(6, N_z, N_x, N_y)$, where $N_x$, $N_y$, and $N_z$ can vary for each atmosphere.

We need to break down the populations into individual pairs of input $X_i$'s and target $y_i$'s, and then group these into training and validation sets. The simplest approach is to consider the 1.5D problem. Each column in the LTE atmosphere is treated as an independent input $X_i$ with shape $(6, N_z, 1, 1)$ and its corresponding column from the NLTE atmosphere is the target point $y_i$ with shape $(6,N_z,1,1)$. Building our training pairs in this way ignores all oblique radiation and therefore the network does not consider the problem in 3D.

By choosing a window of neighboring LTE columns as $X_i$ and the NLTE column corresponding to the middle column of the LTE bundle as $y_i$, we can force the network to consider the problem in 3D. The SunnyNet framework has networks built to handle inputs of size $(6,N_z,1,1)$, $(6,N_z,3,3)$, $(6,N_z,5,5)$, and $(6,N_z,7,7)$. Figure \ref{3d conv} shows an example of a $3\times3$ data pair, with the red LTE pixel being the pixel of interest. The window size is a user-determined variable and should be chosen with consideration to the spatial resolution of the simulation.

After splitting up our data into $(6, N_z, N_x, N_y)$ training pairs, we need to standardize the $z$ dimension, defined as height in the \texttt{BIFROST} simulations. Neural networks are rigid with respect to the size of the input data they can handle, but simulations can have varying height scales and different $N_z$. To solve this problem and make SunnyNet general for simulations with different height scales, we converted the $z$ dimension from a height scale to a column mass scale with a fixed number of points. The column mass is a more relevant quantity for radiative transfer, and the range of column masses that a given spectral line is sensitive to is a more tightly defined quantity than the range of heights, which depend on the particular stratification. 

To convert the populations from height to column mass, we started by computing the average column mass for each height in the simulation, and then interpolated the populations from this scale to our chosen column mass scale. For our runs, we used 400 points for the new column mass scale, evenly spaced on a $\log_{10}$ scale ranging from $10^{-6}$ to $10^2$ kg~m$^{-2}$, which covers the regions in the atmosphere that the hydrogen lines in our model atom are most sensitive to. We did this for both the LTE inputs and the NLTE targets, giving all the data we used a uniform dimension of $(6,400, N_x, N_y)$.

Finally, before feeding the populations to SunnyNet, both in the training and testing, we took the $\log_{10}$ of the populations. This is for two reasons. First, to better condition the problem since the populations for a given simulation column can span more than 17 orders of magnitude. Second, the logarithm of populations will ensure that the predicted populations are always positive and avoid unphysical solutions. The choice of working in log space will have some consequences, which we discuss later.

\begin{figure}
    \centering
    \includegraphics[scale = .25]{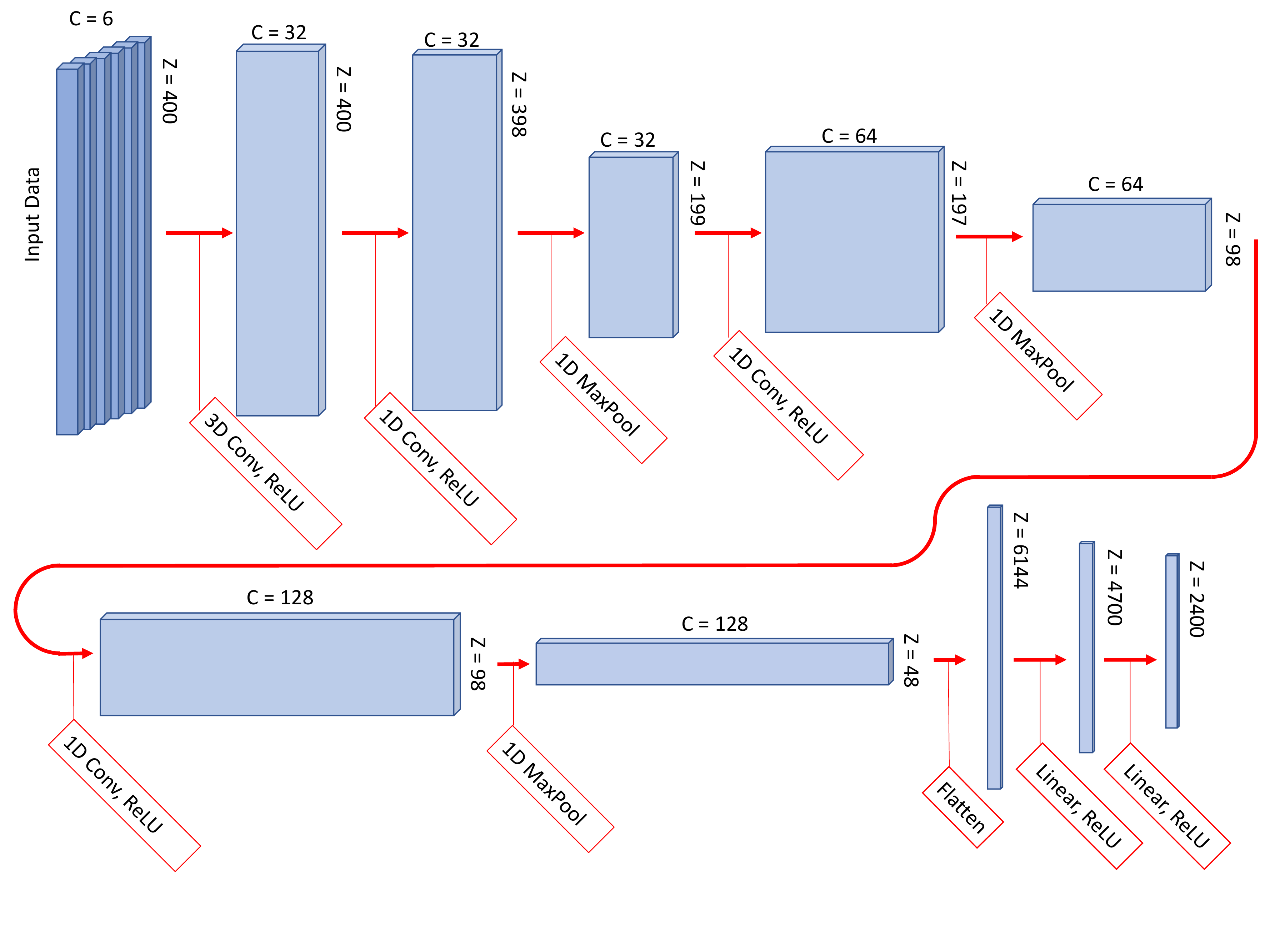}
    \caption{Visualization of one $(6,400,x,y)$ data point passing through the SunnyNet.}
    \label{network picture}
\end{figure}

\subsection{Network structure}

The arrangement of layers often requires quite a bit of guess work, as each application of neural networks is so unique and there is a limited ``best practice'' standard. There are undoubtedly many different and complex network structures that would work well for radiative transfer problems. We found that the architecture shown in Figure \ref{network picture} both performed well and was relatively simple. 

This structure starts with a 3D convolutional layer which can be selected to fit input data with window sizes of $1\times1$, $3\times3$, $5\times5$, and $7\times7$. This is the layer responsible for capturing all of the 3D information across all channels of the input. Figure \ref{3d conv} shows how the 3D convolutional layer processes the input data for the $3\times3$ case. The $(6,3,3,3)$ set of filters (yellow) is restricted to only moving down the data column, pulling out one output value (green) for each convolutional step. Therefore, at each step, the network is gathering information from the point of interests nearest neighbors in all directions. The process is the same for the other input shapes, with the filters taking shapes $(6,3,1,1)$, $(6,3,5,5)$, and $(6,3,7,7)$.

Next come three standard convolutional blocks consisting of a 1D convolutional layer, an activation function, and a 1D MaxPool layer. The Rectified Linear Unit (ReLU) activation function introduces nonlinearity to our network and the MaxPool layer reduces dimensionality in the physical dimension. At the end of our last convolutional block, we flattened all of the learned features and untangle them using two linear layers. We included a drop-out layer between the two linear layers which randomly ``turns off'' nodes and their connections in a layer at a given probability $p$ during each training iteration. This helps to prevent over-fitting by training with a slightly different view of model at each iteration, thus increasing generalization. Our output is a vector of length $2400$, which we then reshape to $(6,400,1,1)$ to match the dimensionality of the target $y_i$ point. 

Table \ref{network table} gives a more in-depth description of the network layers specifically for the $3\times3$ model case. The layers are the same for the other architectures, with the exception of the kernel size of the 3D Conv layer. We followed the convention of PyTorch and grouped the weights and biases into the Learnable Parameters group.

After the data makes its forward pass through the model, we calculated the loss using the following modified mean squared error equation:

\begin{equation}
\begin{split}
Loss & =\frac{1}{N}  (1-\alpha) \sum_{i=1}^{N} \left( \text{NLTE}_{\text{true}_i} - \text{NLTE}_{\text{pred}_i} \right)^2 \\
& + \frac{1}{N} \alpha  \sum_{i=1}^{N} \left( \text{H}_{\text{true}_i} - \text{H}_{\text{pred}_i} \right)^2 .
\end{split}
\end{equation}

This loss compares the predicted and true NLTE energy level populations on a pixel by pixel basis, and also enforces particle conservation. The particle conservation term compares the total hydrogen atom populations at each point in the LTE column of interest ($\text{H}_{\text{true}_i}$) to the total hydrogen atom populations at each point in the corresponding NLTE prediction ($\text{H}_{\text{pred}_i}$). $\alpha$ is a user defined parameter which determines how much each term in the loss function contributes to the total loss.

\begin{table}
\caption{Layers of SunnyNet}
\centering
\renewcommand{\arraystretch}{1.8}
\begin{tabular}{lllr}
    \hline
    Layer Type& \vtop{\hbox{\strut Kernel}\hbox{\strut Shape}}& 
    Output Shape&  \vtop{\hbox{\strut Learnable}\hbox{\strut Parameters}}  \\
    \hline\hline
    3D Conv    & (3,3,3) & $(-1, 32, 400, 1, 1)$ & $5\,216$          \\
    1D Conv    & (3)     & $(-1, 32, 398)$       & $3\,104$          \\
    1D MaxPool & (2)     & $(-1, 32, 199)$       & $0$              \\
    1D Conv    & (3)     & $(-1, 64, 197)$       & $6\,208$          \\
    1D MaxPool & (2)     & $(-1, 64, 98)$        & $0$              \\
    1D Conv    & (3)     & $(-1, 128, 96)$       & $24\,704$         \\
    1D MaxPool & (2)     & $(-1, 128, 48)$       & $0$              \\
    Linear     & --      & $(-1, 4,700)$         & $28\,881\,500$     \\
    Dropout    & --      & $(-1, 4,700)$        & $0$              \\
    Linear     & --      & $(-1, 2,400)$         & $11\,282\,400$     \\
    \hline\hline
\end{tabular}
\renewcommand{\arraystretch}{1.2}
\begin{tabular}{p{.88\textwidth}}
    Input shape: $(-1, 6, 400, 1, 1)$  \\
    Total Parameters: $40\,203\,132$         \\
    Parameter size: 153.36~MB      \\
\end{tabular}
\tablefoot{
The layers are shown in order, from the first 3D convolutional layer to the output linear layer. The stride for 3D Conv, 1D Conv, and 1D MaxPool are one, one, and two respectively. The 3D Conv layer is also 0-padded in the $z$ direction. The $-1$ dimension in the ``Output Shape'' column is a placeholder for however many input samples are in each training batch.
}
\renewcommand{\arraystretch}{1.8}
\label{network table}
\end{table}

\begin{table}
\caption{Hyperparameters used}
\centering
\renewcommand{\arraystretch}{1.2}
\begin{tabular}{l c}
    \hline\hline
    Epochs         & 50                 \\
    Batch Size     & 128                \\
    Optimizer      & Adam\tablefootmark{a}               \\
    Learning Rate  & $10^{-3}$ \\
    $\alpha$       & $10^{-3}$\tablefootmark{b}                \\
    Early Stopping & 5                  \\
    \hline\hline
\end{tabular}
\tablefoot{
\tablefoottext{a}{Refers to the Adam optimizer of \cite{adam}.} 
\tablefoottext{b}{$\alpha$ is the scaling factor used in the loss function.} 
}
\label{hyperparams}
\end{table}

\subsection{Spectral synthesis}

After running the neural network, the output will be the predicted NLTE populations for our model atom for every point in the 3D grid. The final step is to solve equation (\ref{eq:rte}) and obtain $I_\lambda$ for the spectral line(s) and direction(s) of interest. To do this, we used the atomic level populations and 3D atmosphere to compute the total source function $S_\lambda$. For a given wavelength, $S_\lambda$ is defined as the ratio of the emissivity $j_\lambda$ by the extinction coefficient $\alpha_\lambda$, which comprise both continuum and line contributions:
\begin{equation}
S_\lambda \equiv \frac{j_\lambda}{\alpha_\lambda} = \frac{j_{\lambda}^{\mathrm{cont}}+ j_{\lambda}^{\mathrm{line}}}{\alpha_{\lambda}^{\mathrm{cont}}+\alpha_{\lambda}^{\mathrm{line}}}.
\end{equation}
We made use of the \texttt{Transparency.jl} library \citep{transparency} to compute the line and continuum emissivities and perform the formal integration of equation (\ref{eq:rte}). As continuum sources we included free-free extinction from H$^-$, H$^+_2$ molecules and hydrogenic species, and bound-free extinction from H$^-$ and H$^+_2$ molecules, together with Thomson scattering and Rayleigh scattering from H atoms. For each spectral line, we used
\begin{eqnarray}
\alpha_{\lambda}^{\mathrm{line}} &=& \frac{hc}{4\pi \lambda_0}\left(n_l B_{lu} - n_u B_{ul} \right) \varphi(\lambda-\lambda_0), \\
j_{\lambda}^{\mathrm{line}} &=& \frac{hc}{4\pi \lambda_0} n_u A_{ul} \varphi(\lambda-\lambda_0),
\end{eqnarray}
where $n_l$ and $n_u$ are the populations of the lower and upper levels, $A_{ul}$ are the Einstein coefficients for spontaneous de-excitation, $B_{lu}$ and $B_{ul}$ the Einstein coefficients for stimulated excitation and de-excitation, and $\varphi(\lambda-\lambda_0)$ the line profile. Here we assumed the same line profile for extinction and emission, and therefore are under the approximation of complete redistribution (CRD).

For all the spectral synthesis in this work we limited our calculations to the H$\alpha$ line, since it is a strong line that is influenced by 3D effects \citep{Leenaarts:2012} and a widely used diagnostic of the solar chromosphere. The line is most sensitive the populations of its upper and lower levels ($n=3$ and $n=2$), but it is also sensitive to the ground-level population of hydrogen ($n=1$) and of ionized hydrogen ($n=6$ in our model atom) through the continuum contribution and total number of absorbers. For simplicity we also limited our analysis to line profiles for emergent intensity along the vertical direction (i.e., solar disk-center intensity), but the method is general for any viewing angle.

\begin{figure*}
    \centering
    \includegraphics{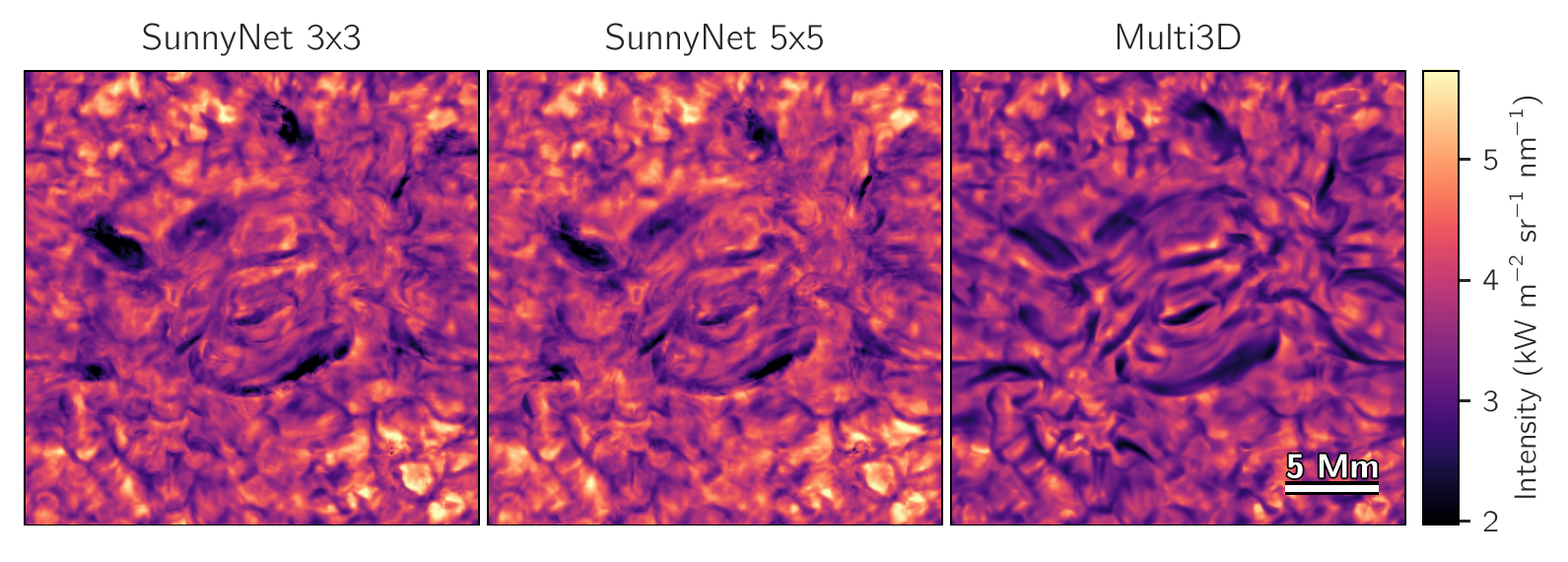}
    \caption{H$\alpha$ line center intensity for the enhanced network simulation at $t=4250$s using populations from Multi3D and SunnyNet with $3\times3$ and $5\times 5$ window sizes.}
    \label{3 vs 5 maps}%
\end{figure*}

\section{Results}

\subsection{Running time}

The appeal of machine learning techniques to solve the 3D NLTE problem is in great measure to save computational time. Hence, it is of great interest to see how SunnyNet performs in this regard. We ran all our tests on a single machine, with two AMD EPYC2 7302 3 GHz CPUs (each 16-core) and one NVIDIA Tesla T4 GPU. PyTorch does the heavier computations in the GPU, but some parts are also multithreaded in the CPU cores. Excluding the calculation of synthetic H$\alpha$ intensity, which is done via a separate script, SunnyNet performs three types of tasks: preparing input files, training the network, and predicting the populations. The first involves reading the 3D models and preparing them into a form that SunnyNet can read. These tasks take usually about one minute of wall time; the code is serial and most of the time is spent on I/O operations. The training and predicting parts is mostly run on the GPU. 

For a simulation with $252\times252\times460$ grid points, training took about one minute per epoch. For our cases the training took about 15-30 epochs, meaning a wall time between 15 to 30 min. The prediction step was much faster, since it only took one forward pass of the data through the model, and typically took between 30~s to one minute (most of the time spent in serial I/O and rearranging the arrays). The time for the prediction step is the one we should compare with typical 3D NLTE codes. Running Multi3D for the same setup took between $20\, 000$ and $100\, 000$ CPU hours, depending on the CPU model and simulation snapshot, and was typically run using several thousand CPU cores over more than a day. Therefore, the SunnyNet speedup is about $10^5$ times compared to single CPU core performance. Finding the equivalence between one GPU plus some multithreaded CPU code with a single-CPU code is not straightforward, but this is a conservative estimate.

Computing the H$\alpha$ intensity with our Julia script (making use of 32 CPU threads) for 101 wavelength points and a single viewing angle took about one minute for $252\times252\times460$ grid points and 12 min for $720\times720\times635$ grid points.

\subsection{Window size}

We started our analysis by finding out the best-performing input parameters for SunnyNet. Then we proceeded to test SunnyNet in increasingly challenging problems. First, we trained and test SunnyNet using different snapshots from the same simulation. Afterwards, we used different simulations for testing and training, looking at simulations with different magnetic field configurations and also different spatial resolutions. These tests will give us insight into how generalized the models are across different simulations.  

All models in our analysis have an identical structure except for the first 3D Conv layer. They are trained with the same loss function, and use the same training hyperparameters given in Table \ref{hyperparams}. The parameters we used work well for all tested simulations and were found using best practices and some trial and error. The intention in not conducting a meticulous hyperparameter search is generalization. Endless hours could be spent fine-tuning a model to perform well on a specific simulation, but this could result in over-fitting to the limited data set, and loss of generalization. We chose, instead, to focus on testing our architecture's performance across various atmospheric conditions using set hyperparameters.

\begin{figure}
    \centering
    \includegraphics[width=0.49\textwidth]{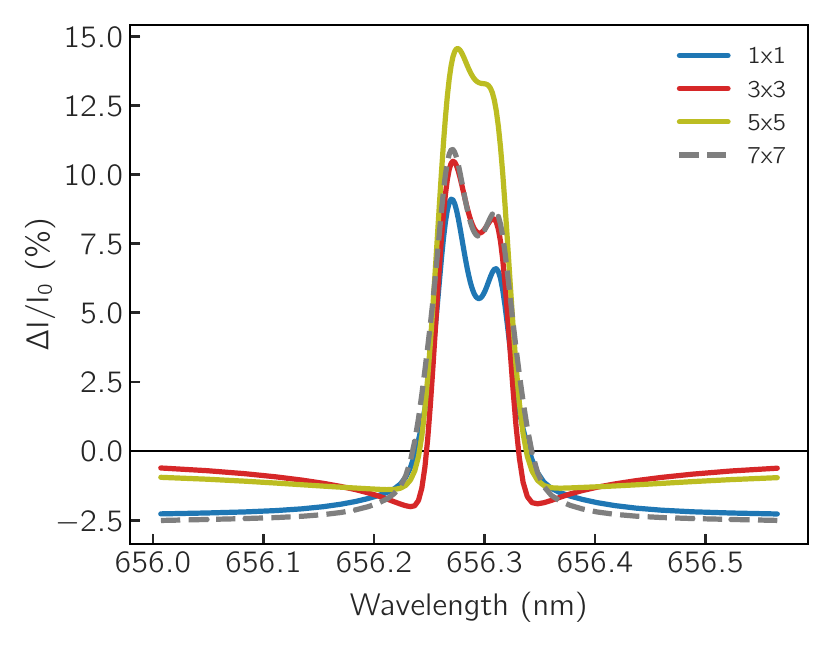}
    \caption{Relative error of H$\alpha$ mean intensity (compared to intensity from Multi3D populations) for the enhanced network simulation, snapshot at $t=4250$~s.}
    \label{avg intensity}
\end{figure}

We tested the base SunnyNet architecture with four different input window sizes of $1\times1$, $3\times3$, $5\times5$, and $7\times7$. The training set for the models was built from two snapshots from the enhanced network simulation (described in §\ref{sec:qs}), with a horizontal pixel size of $96\, \mathrm{km} \, \mathrm{pix}^{-1}$. The models were tested on two different snapshots from the same simulation as the training data. For both of our test snapshots, the predictions from the $3\times3$ models give the lowest errors.

In Figure~\ref{3 vs 5 maps} we compare the synthetic H$\alpha$ line core intensity using populations from SunnyNet with different window sizes and populations from Multi3D, and in Figure~\ref{avg intensity} compare the effect of window sizes for all H$\alpha$ wavelengths of the horizontally-averaged mean spectra, relative to $I_0$, the intensity using populations from Multi3D. The results shown are for one snapshot, but results for the snapshot at $t=4890$~s (not shown) are very similar -- that snapshot shows overall intensities closer to the ones from Multi3D, but the differential effect of window sizes shows the same behavior. The figures illustrate that the window size that gives the lowest population errors also gives the best agreement with the synthetic intensities in H$\alpha$, although the differences between some window sizes are small. Since the window sizes are defined in pixels and not in a physical quantity such as km, the best fitting size will vary for different simulations. The key relation here is how the pixel size relates to the photon mean free path $l_\lambda$, since $l_\lambda$ defines how many grid points a typical photon can travel, and is the relevant quantity for assessing the importance of 3D effects. For higher resolution simulations one will need to use larger window sizes, as a typical $l_\lambda$ is covered by more grid points. %

Using larger window sizes allows for the network to consider a larger 3D box around the target column. However, if $l_\lambda \lesssim 1\,\mathrm{pix}$, increasing the window size will lead to the network being fed more irrelevant data and possible over-fitting. The relation does not seem very clear cut, with $5\times5$ performing worse near the line core (longer $l_\lambda$) and better at the wings (shorter $l_\lambda$), while $7\times7$ does better at the line core but worse in the wings. Since a window size of $3\times3$ gives the best results, we henceforth use it for the rest of our analysis.

\subsection{Enhanced network simulation\label{sec:qs}}

\begin{figure}
    \centering
    \includegraphics{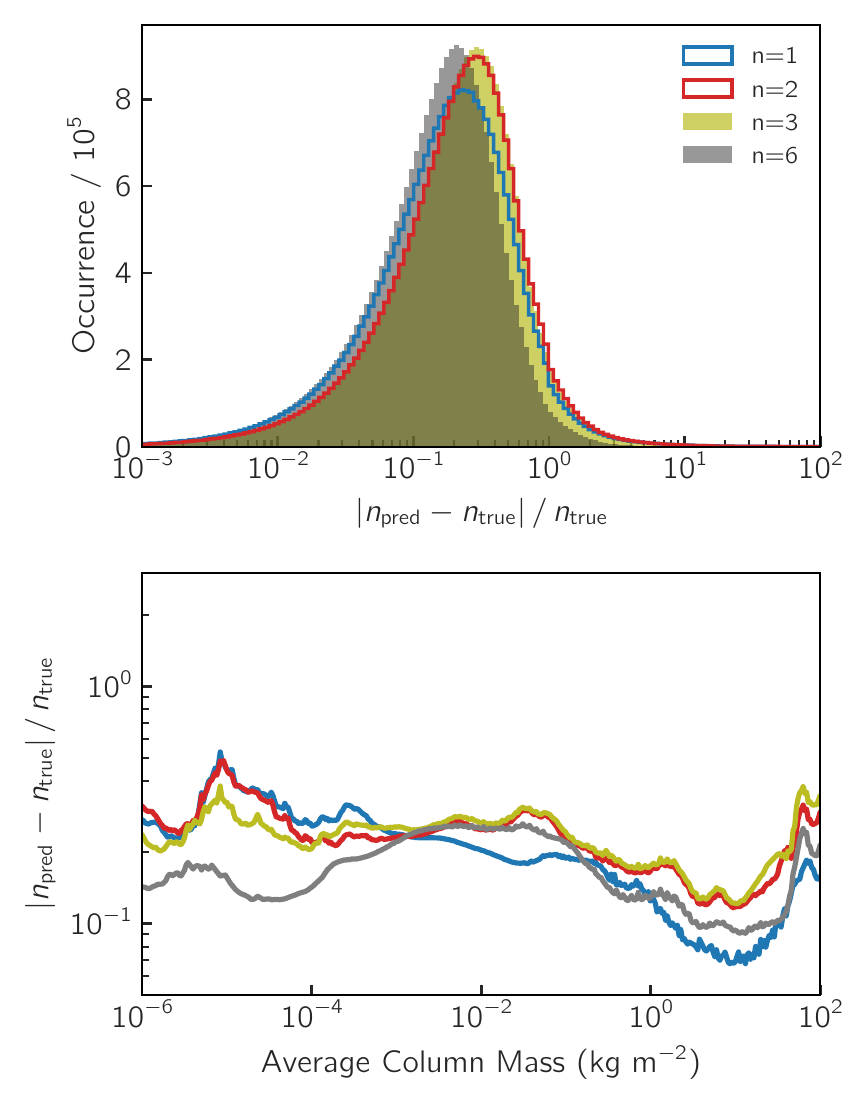}%
    \caption{Normalized departures between predicted and true populations for the enhanced network simulation ($t=4890$~s), for four levels of hydrogen ($n=6$ represents H II). \emph{Top:} histograms for all points in the simulation. \emph{Bottom:} median of the departures in the horizontal direction, as a function of average column mass. The colors for different levels are the same for both panels.}
    \label{cb24bih_489_hist}
\end{figure}

The first model we analyzed was trained on hydrogen populations calculated from the simulation by \citet{cb24bih}. This is a simulation of a ``enhanced network'' quiet Sun, and accounts for nonequilibrium hydrogen ionization. The simulation box has a physical size of $24 \times 24 \times 15.4$ Mm$^3$ and horizontal resolution 48~$\mathrm{km}\ \mathrm{pix}^{-1}$. For all our calculations we did not use the full resolution, but sampled every other spatial pixel, effectively halving the resolution to 96~$\mathrm{km}\ \mathrm{pix}^{-1}$, with $252\times 252\times 460$ pixels used in the Multi3D run. The magnetic field setup consists of two regions of opposite polarities separated by 8 Mm that are injected at the bottom boundary as vertical field. The mean unsigned magnetic field strength is 4.8~mT (48~G) at the photosphere. The training set was constructed using the populations calculated from two snapshots at $t = 3850$~s and $t=5300$~s respectively. We then used the populations from the two intermediate snapshots, at $t = 4250$~s and $t=4890$~s, as test sets for the network. While we analyzed both snapshots, in the interest of brevity, all the figures shown henceforth are for the snapshot at $t=4890$~s. There are only small differences in results for the different test snapshots, with SunnyNet performing slightly worse in the line core for the snapshot at $t = 4250$~s.

We first looked into how the predicted hydrogen populations from the neural network compare to the populations obtained with Multi3D. The plots in Figure \ref{cb24bih_489_hist} show histograms for the absolute value of the relative differences between the true and predicted populations for all points in the simulation grid, together with a plot of the median of the same quantity, but taken along horizontal slices for each value of the vertical scale (in average column mass). The histograms reflect how well SunnyNet estimates the populations of different levels in the whole box, while the second quantity gives an overview of any differences in the quality of estimation with height. In the regions with higher column mass (deeper in the atmosphere), the populations are very close to LTE (i.e., the original estimate that was fed to SunnyNet), while for outer layers the populations are several orders of magnitude away from the LTE values (below an average column mass of $10^{-3}$~kg~m$^{-2}$, $n_\mathrm{NLTE} / n_\mathrm{LTE}$ can be up to $10^{18}$, although median values are around $10^{3}-10^{7}$ depending on the level). This means that at lower column masses, SunnyNet's predicted values depart strongly from its inputs, correctly following the true result.

The results show that for the enhanced network simulation, SunnyNet is able to predict the NLTE populations for most points within 20-30\%. The median normalized difference for all levels is 0.204, the 10th percentile is 0.036 and the 90th percentile is 0.634. The quality of the prediction is nearly constant across different levels -- the ground ($n=1$) and ionized ($n=6$) are slightly better predicted than the other excited states, but the differences are small. The performance of SunnyNet also shows little variation with height. In Figure~\ref{cb24bih_489_hist}, bottom panel, one sees that the median of the distribution changes from approximately 0.1 to 0.35 and again all levels seem to follow the same trend. %

\begin{figure*}
    \centering
    \includegraphics{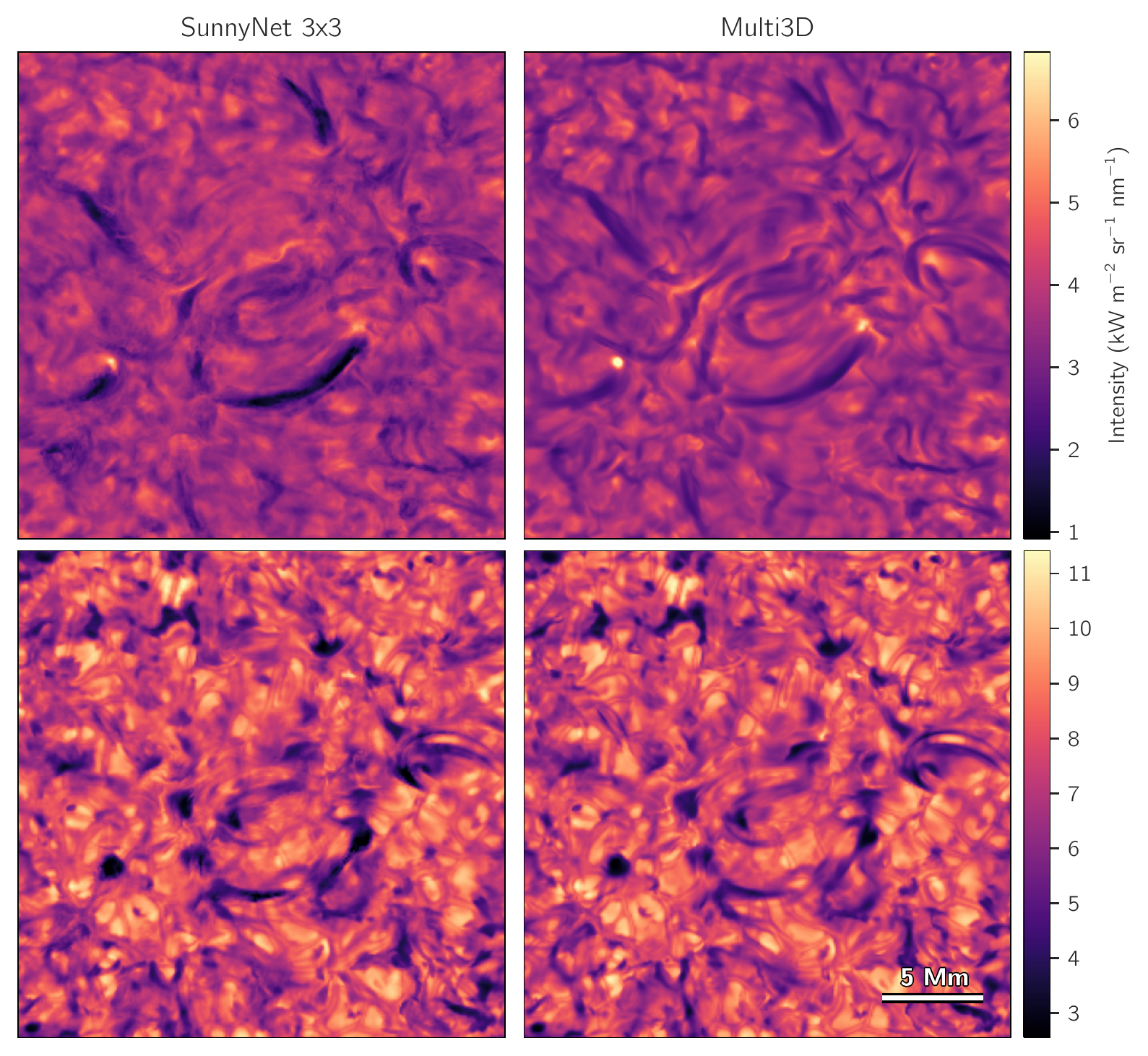}
    \caption{H$\alpha$ intensity for the enhanced network simulation at snapshot $t=4890$s for the line center (top) and red wing at $v = 15.96$ km~s$^{-1}$ (bottom).}
    \label{cb24bih_489}
\end{figure*}

\begin{figure}
    \centering
    \includegraphics{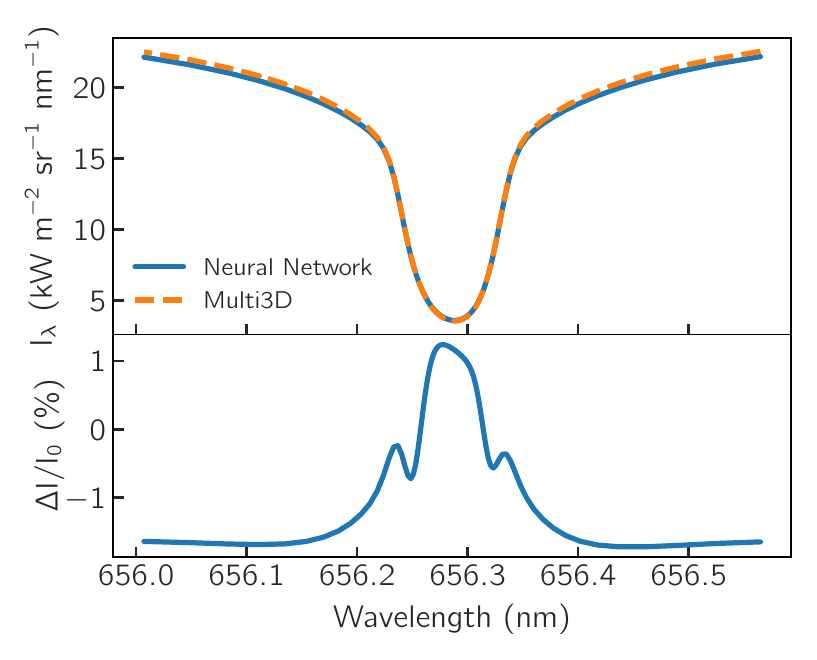}
    \caption{Spatially averaged H$\alpha$ spectra for the enhanced network simulation snapshot at $t=4890$~s, covering the wavelength range of $\pm 127\; \mathrm{km}\;\mathrm{s}^{-1}$ around the line core.}
    \label{cb24bih_diff}
\end{figure}

One could expect that at deeper layers, where LTE conditions dominate and the populations depart very little from the initial input, that the predictions from SunnyNet would be much more accurate that at upper layers, where the true result is orders of magnitude off the initial input. However, this is a reflection on how SunnyNet treats all layers equally. The network is not informed about the physics of when LTE is valid, and tries to find the best fitting solution also allowing for changes in populations in regions where LTE dominates. This is also a consequence of SunnyNet working in the log space of populations. In deeper layers the populations are many orders of magnitude higher than in the outer layers, and so a relatively small difference, for example, 0.1~dex will result in a very large absolute difference (and a less physical solution) in the deeper layers, than when compared to the same small relative difference in the outer layers. 

We looked also at the quality of the predictions versus the departure coefficients $b\equiv n_\mathrm{NLTE} / n_\mathrm{LTE}$, to see if SunnyNet's performance was being affected when the true result is departs more strongly from the initial input. The comparison shows that it is not. The distribution of normalized differences vs departure coefficients is flat for almost all levels, the only exception is that the ground level does slightly worse when $b>10^{7}$, but all the other levels show a flat distribution (and even a very slight improvement) as $b$ increases.

In Figure \ref{cb24bih_489} we compare H$\alpha$ synthetic intensity maps computed with the populations predicted from SunnyNet and with the populations computed by Multi3D. We look at two wavelengths: the line center and a position on the red wing at 15.96~$\mathrm{km}\ \mathrm{s}^{-1}$ from the line center. In both cases the overall morphology is well reproduced but there are some localized differences. This is a pattern that repeats itself when analyzing different snapshots (see Figure~\ref{3 vs 5 maps} for a comparison of the line center for the $t=4250\ s$ snapshot).

\begin{figure}
    \centering
    \includegraphics{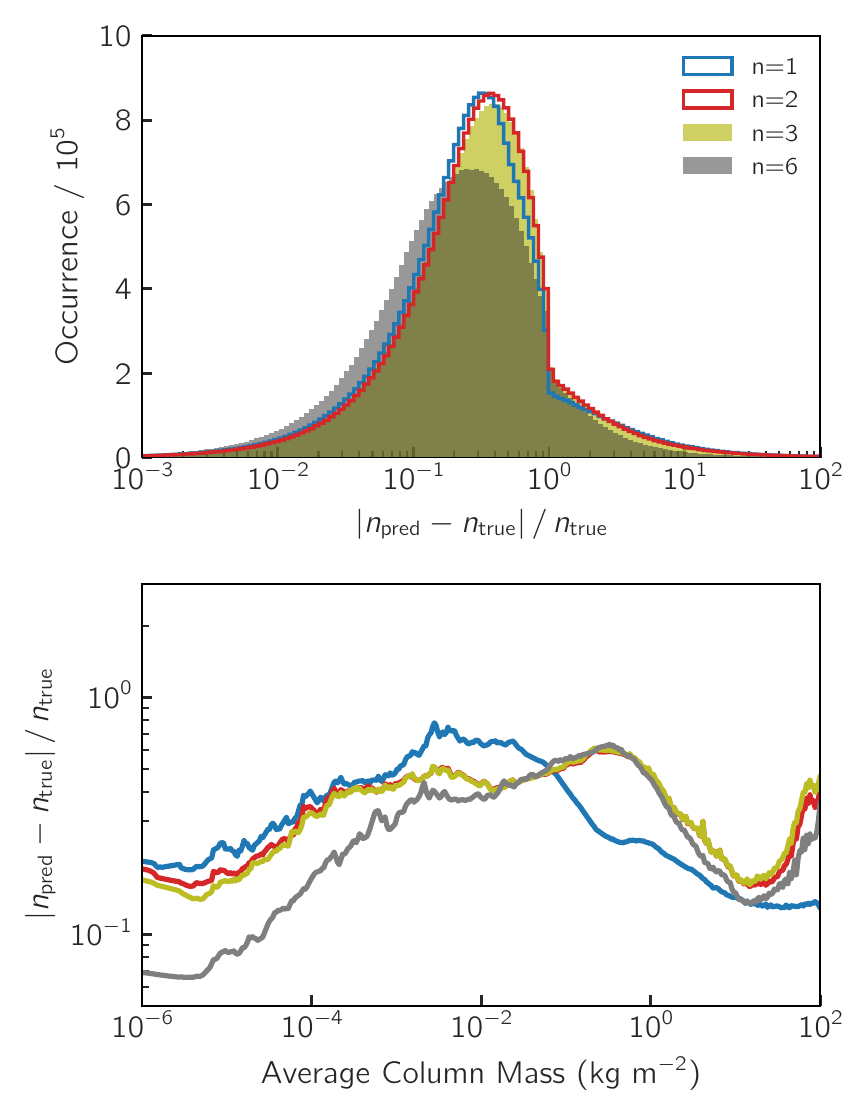}%
    \caption{Same as Figure~\ref{cb24bih_489_hist} but for the flaring sun simulation at $t=9570$s.}
    \label{cbh24_0957_hist}
\end{figure}

To study how other wavelengths compare on average, we plot in Figure~\ref{cb24bih_diff} the mean spectra computed from the SunnyNet and Multi3D populations, and also the relative difference in intensity. With the populations from SunnyNet, the wing intensity is slightly underestimated, while around the H$\alpha$ core it is overestimated, slightly above the 1\% level. For snapshot $t=4250\ s$ (not shown) the core intensity overestimation is slightly larger, up to 10\%, but again for the wings and most of the profile the difference is much lower. The reason why the line core for the $t=4250\ s$ case is, on average, not as well reproduced as the continuum is not completely clear. The $n=1$ and $n=6$ levels are slightly better predicted than the other two levels, and they influence mostly the continuum intensity. A possible explanation of the difference is because the radiation near the line core is formed over a wider range of heights, and therefore sensitive to a much larger fraction of the predicted populations, which can locally have worse estimates, as seen in Figure~\ref{cb24bih_489} where the differences in spatial structures are larger at the line core than at the wing.

The synthetic intensities show that, while the population estimates are accurate to about 20--30\%, the effect of this difference in the intensities is relatively small. Both intensity maps and mean spectra are much closer than the difference in populations, since the radiative transfer equation is not linearly sensitive to the populations.

\subsection{Flaring simulation}

To test SunnyNet in a wider range of physical properties, we tested it also using the Bifrost simulation described in \citet{cbh24}. This solar simulation is more active: a horizontal flux sheet of 336 mT (3.36 kG) was injected, and as a consequence, the simulation developed several small flaring events, which are covered in the snapshots we studied. 

The physical size of this simulation is the same as the enhanced network simulation and again we ran the radiative transfer calculations on every second pixel in the horizontal scale, resulting in a horizontal resolution of 96~$\mathrm{km}\ \mathrm{pix}^{-1}$ and $252\times 252\times 467$ pixels used in the Multi3D run. We followed the same procedure as before, creating the training set using the populations from two snapshots of the simulation. These snapshots are at $t=9100$~s and $9590$~s. As before, we tested SunnyNet using multiple snapshots but show in figures only the results for the snapshot at $t=9570$~s -- there were no significant differences in the results for different snapshots.

\begin{figure*}
    \centering
    \includegraphics{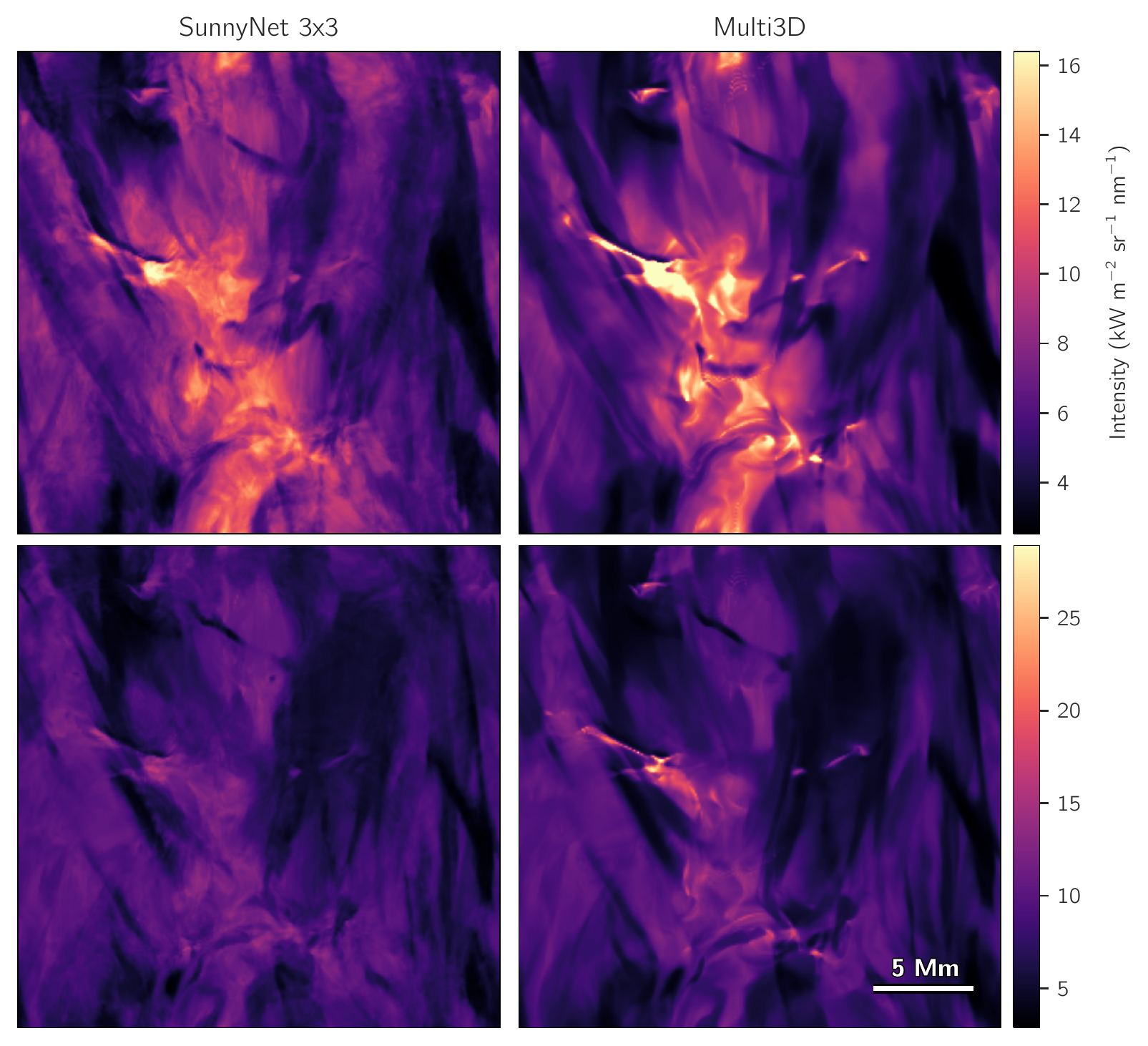}
    \caption{H$\alpha$ intensity for the flaring simulation at snapshot $t=9570$s for the line core (top) and red wing at  $v = 15.96$ km~s$^{-1}$ (bottom).}
    \label{cbh24_0957}
\end{figure*}

\begin{figure}
    \centering
    \includegraphics{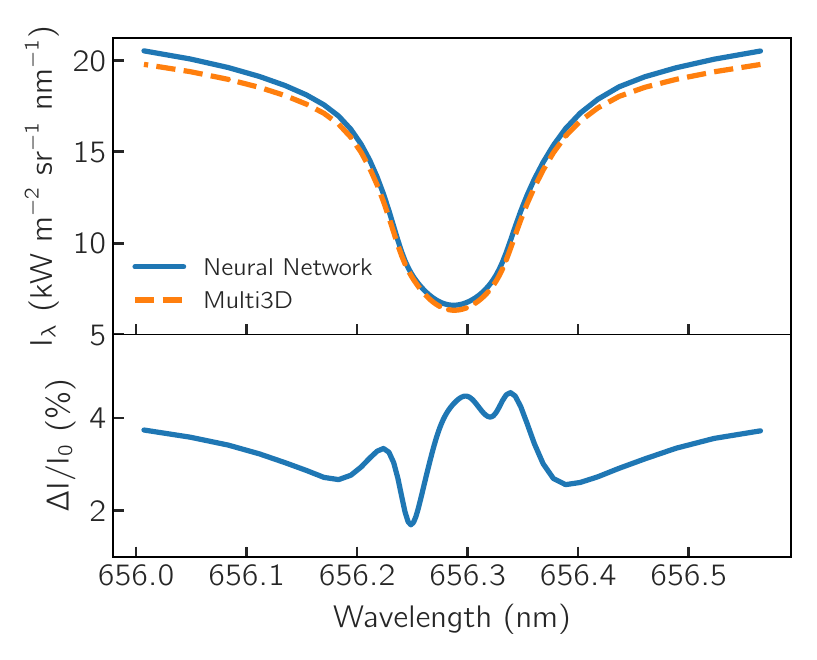}%
    \caption{Spatially averaged H$\alpha$ spectra for the flaring simulation snapshot at $t=9570$~s, covering the wavelength range of $\pm 127\; \mathrm{km}\;\mathrm{s}^{-1}$ around the line core.}
    \label{cbh24_diff}
\end{figure}

Figure \ref{cbh24_0957_hist} shows the histograms and median differences vs mean column mass for SunnyNet's predictions for the $t=9570$~s snapshot. The results are similar to those of the enhanced network. Overall, SunnyNet is able to predict the NLTE populations for most points within 30--40\%. The median normalized difference for all levels is 0.286, the 10th percentile is 0.049 and the 90th percentile is 1.015. This is not as good as the prediction for the enhanced network simulation, but comparable. While the median values are higher, the histogram distribution is also not as symmetric when plotted against the logarithm of the normalized differences -- there is a marked decrease after a difference of 100\%. The quality of the predictions vs average column mass also varies more than for the enhanced network simulation: for some reason the predictions are not as good around an average column mass of $10^{-2}~$kg~m$^{-2}$, and improve for larger and smaller column masses. The $n=2$ and $n=3$ predictions are almost equally good and show nearly the same profile with average column mass. For $n=1$ the predictions are a little better, in particular for deeper layers, while $n=6$ looks on average about as good, but more accurate for the outer layers. 

The flaring simulation is much more dynamic and contains more extreme conditions than the enhanced network simulation. Due to the intermittent flaring there is also a larger snapshot-to-snapshot variation. For these reasons it is perhaps not so surprising that SunnyNet's estimates are not as accurate. In this simulation there is also a wider range of departures from LTE, especially for the ground and ionized states. Again we see little correlation between the quality of the estimates with the departures from LTE, meaning SunnyNet is performing at the same level even in the most challenging regions.

In Figure~\ref{cbh24_0957} we compare synthetic H$\alpha$ images at the line core and wing using populations predicted by SunnyNet with the ones from Multi3D, for the vertical emergent intensity. Again, the overall morphology is generally very well replicated in the SunnyNet maps. The dark filamentary parts are well reproduced, but the predictions are worse for the very bright regions where the flaring is taking place, in particular for the line center intensities. A possible reason for this is that pixels with very energetic phenomena are more rare, and the network focuses more on the most representative solutions. A closer look at the intensity maps shows that SunnyNet maps are not as smooth as the ones from Multi3D, showing some spurious substructure. This subtle effect is more evident in the line center images, but can also be seen in the wing images. This could be a consequence of using a $3\times 3$ window size, while the Multi3D calculations allow for rays spanning the entire box.

In Figure~\ref{cbh24_diff} we compare the spatially averaged H$\alpha$ spectra using populations from SunnyNet and Multi3D. We find that SunnyNet's predictions lead to slightly higher mean intensities across the line profile, about $3-4$\% above those of Multi3D, being more evident in the wings because of their higher intensity. There is also some snapshot to snapshot variation with the $t=9570$~s snapshot giving results closer to Multi3D. This shows that despite an overall worse prediction of the populations, the mean spectra are still reasonably well reproduced.

\subsection{Out-of-sample simulations}

So far we have trained and tested SunnyNet using different snapshots from the same simulation, using it to predict populations from any snapshot in the simulation time series. Our next step is much more demanding: to train and test using different simulations. We constructed a training set built from two snapshots from the enhanced network simulation and two snapshots from the flaring simulation, and tested SunnyNet for two other simulations, described below. Although all simulations were run with the Bifrost code and share similar physics inputs such as the equation of state, and treatment of radiation, they have substantial differences in height stratification, magnetic configuration, and physical size, as described below. The goal with these ``out-of-sample'' tests was to run a worst-case scenario for SunnyNet, using simulations substantially different from those used for the training.

The first out of sample simulation is both larger and much deeper, with a size of $72\times72\times64$~Mm$^3$, with 8.5~Mm above the photosphere. We refer to this simulation as the ``extended'' simulation. The simulation is part of a study of flux emergence by \citet{Hansteen:2020proc}, and was provided courtesy of V. H. Hansteen. It was started with a 10 mT horizontal magnetic field throughout the convection zone, and a sheet with 20 mT was injected at the bottom boundary, followed by injections of 100~mT after 63 min and 200~mT after 133 min, decreasing to 30 mT at 288 min. This flux took a few hours to reach the surface, and the snapshot we used here happens when a good amount of mixing and flux emergence was already underway at the surface, at about 367 min from the start. This extended simulation has a horizontal resolution of 100~$\mathrm{km}\ \mathrm{pix}^{-1}$, which is nearly the same as the pixel size we used for the training simulations, just a much larger spatial extent and number of pixels: $720\times 720 \times 635$, about 8 times larger than the training simulations. 

The second out of sample simulation has a much smaller physical size but higher spatial resolution. We refer to it as the ``high resolution'' simulation. Its spatial extent is $6\times 6\times 10.3$~Mm$^3$, and it is part of an experiment with a higher spatial resolution of 23~$\mathrm{km}\ \mathrm{pix}^{-1}$. This simulation was provided courtesy of M. Carlsson. It is much more quiet than the other simulations, with a mean unsigned magnetic field in the photosphere of about 0.6 mT. Although this simulation has a much smaller spatial extent, it has about the same number of pixels as the training simulations: $256\times 256 \times 430$. Given its reduced spatial coverage and weak magnetic fields, this simulation lacks prominent magnetic loops or long chromospheric fibrils. The aim of including this simulation was not just to test a more quiet Sun configuration, but also to see how SunnyNet would fare when testing a simulation with a much higher spatial resolution than the training simulations. In both training and testing the window size was kept at $3\times 3$. Because the simulations have different resolutions, the three pixels of the window have different physical sizes between the training and testing simulations, which leads to inconsistencies in the spatial extent of the windows. This was intentional: we wanted to test how badly this would affect the predictions.

\begin{figure}
    \centering
    \includegraphics{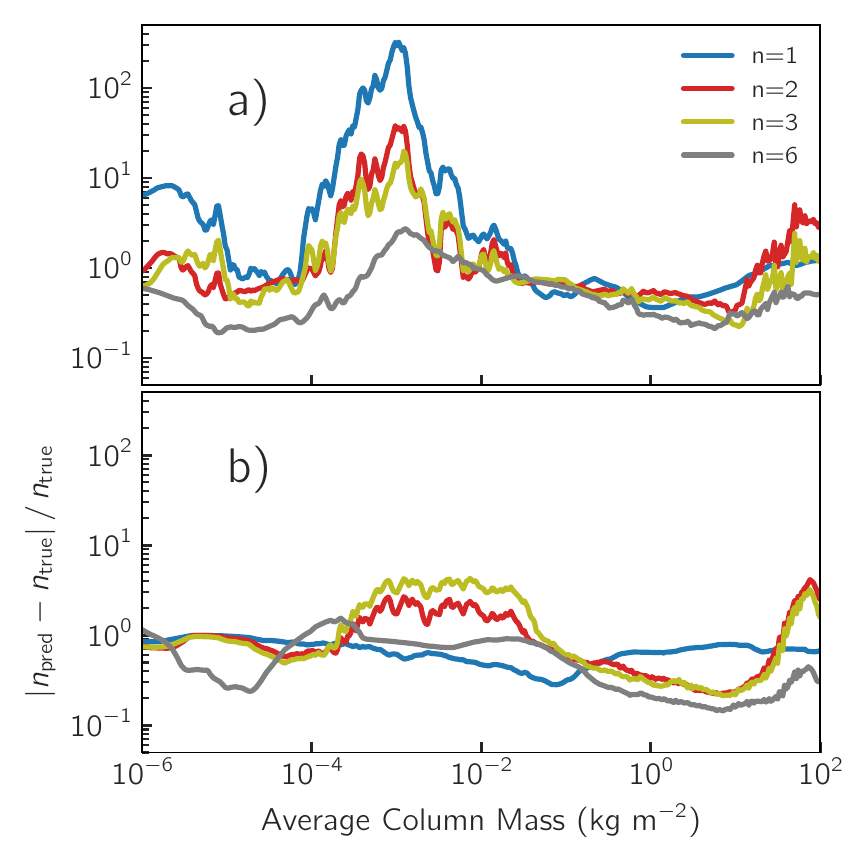}
    \caption{Same as Figure~\ref{cb24bih_489_hist} (bottom panel), but for the out-of-sample simulations. \emph{Top:} extended simulation. \emph{Bottom:} high-resolution simulation.}
    \label{errors_out}
\end{figure}

\begin{figure*}
    \centering
    \includegraphics{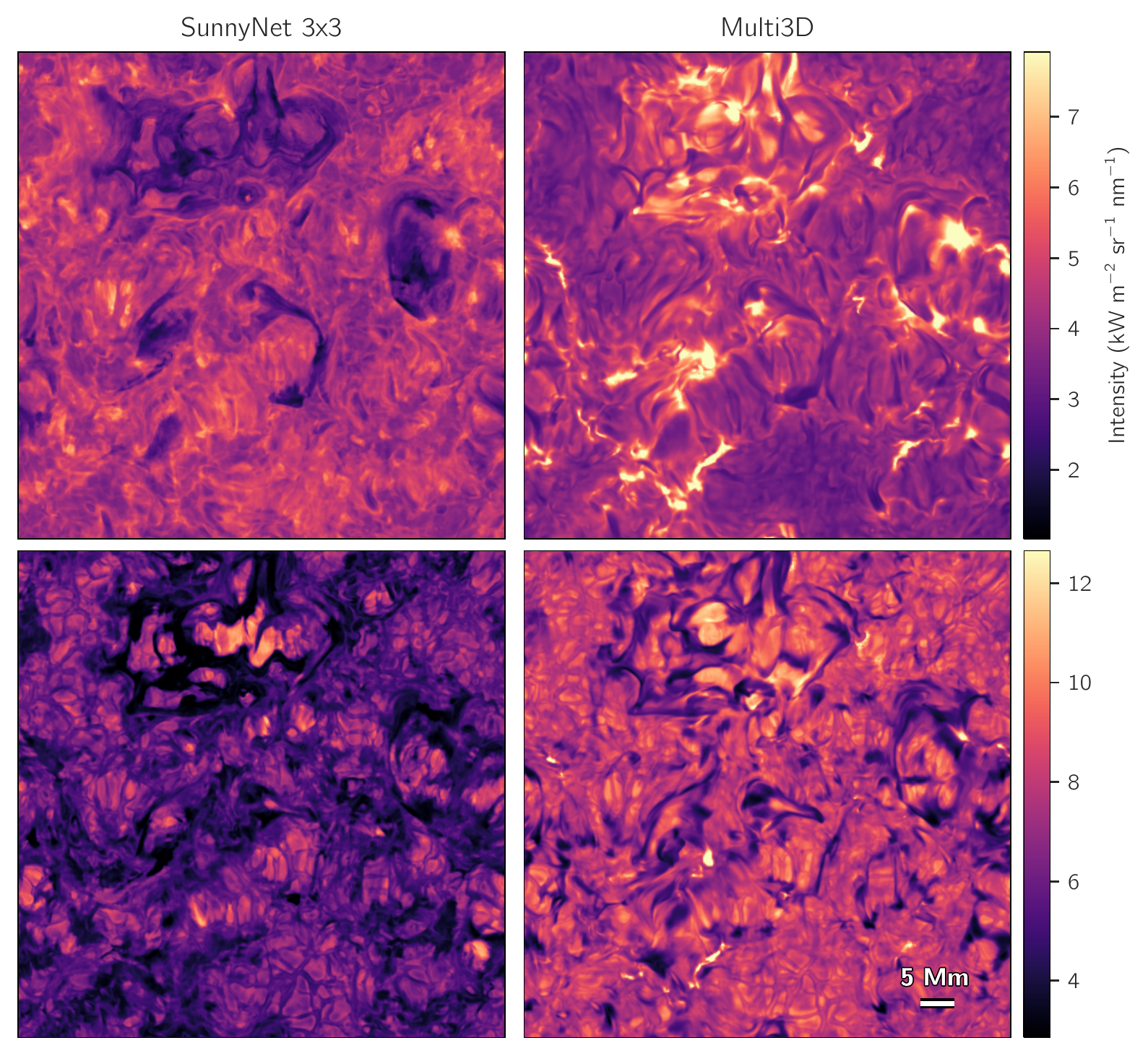}
    \caption{H$\alpha$ intensity for the extended simulation for the line core (top) and red wing at $v = 15.96$ km~s$^{-1}$ (bottom).}
    \label{nw072100}
\end{figure*}

\begin{figure*}
    \centering
    \includegraphics{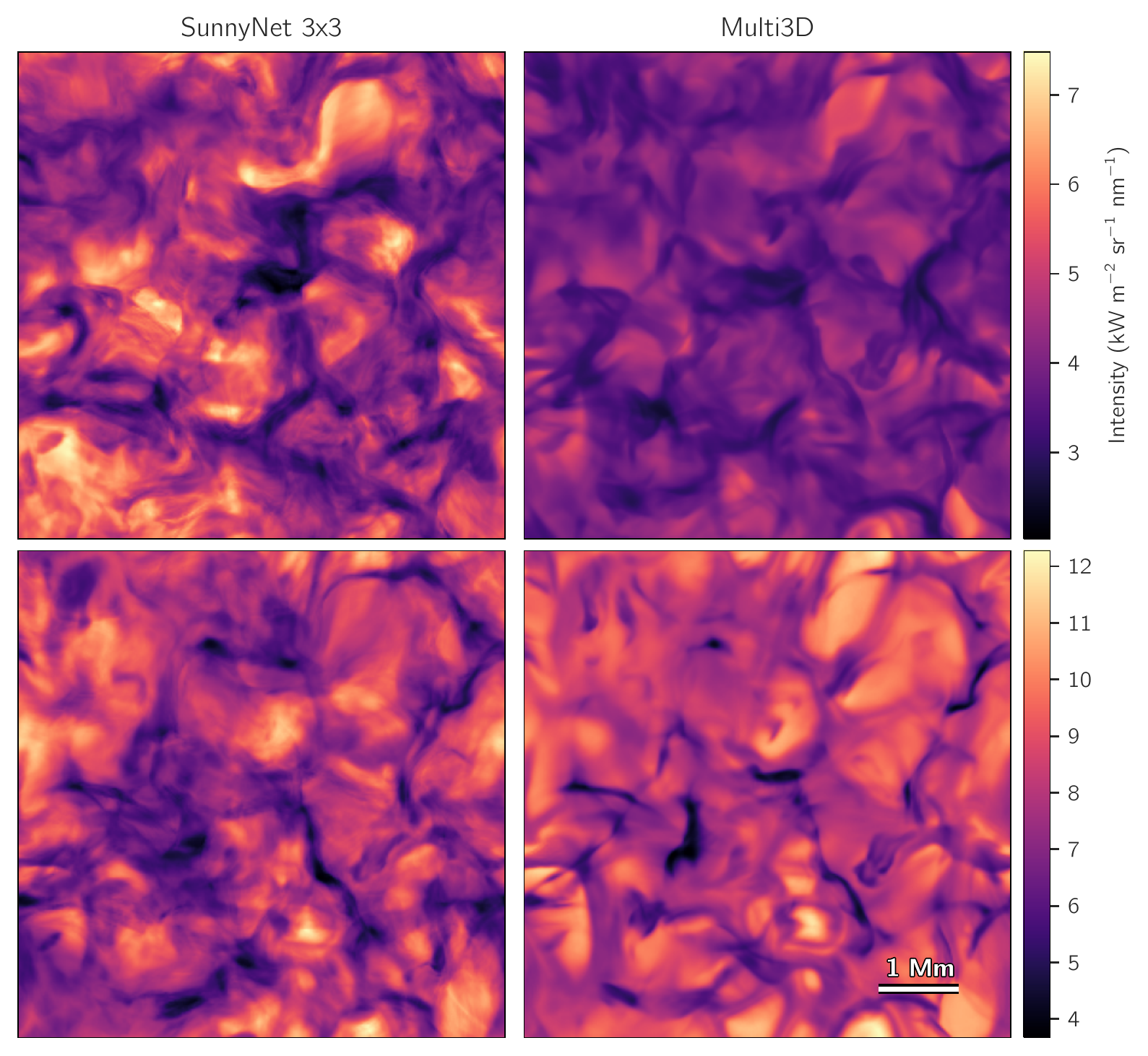}
    \caption{H$\alpha$ intensity for the high-resolution simulation for the line core (top) and red wing at $v = 15.96$ km~s$^{-1}$ (bottom).}
    \label{qs006023}
\end{figure*}

Only one snapshot was tested for each of the out-of-sample simulations. For the sake of brevity only a subset of the analysis is shown in figures. The median normalized differences as functions of average mean column mass (as in the bottom panel of Figure~\ref{cb24bih_489_hist}) are shown for both simulations in Figure~\ref{errors_out}. Here we see a very different behavior from before. For average column masses larger than 1~$\mathrm{kg}\;\mathrm{m}^{-2}$ the median differences are larger than, but comparable, to the results for the enhanced network and flaring simulations. However, for smaller average column masses (layers above the solar surface), the estimates from SunnyNet are orders of magnitude off. 

\begin{figure}
    \centering
    \includegraphics{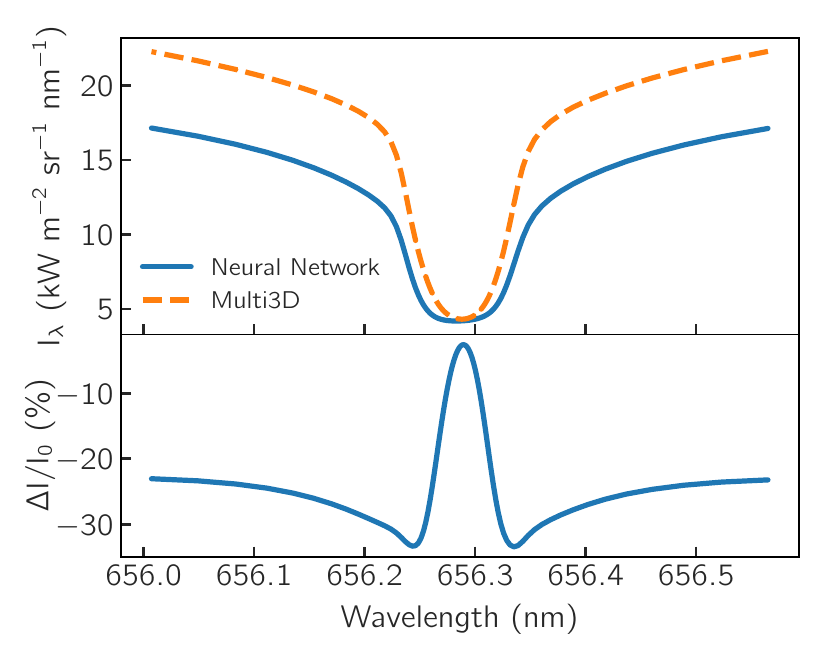}\\
    {\footnotesize (a) Extended simulation}\\
    \includegraphics{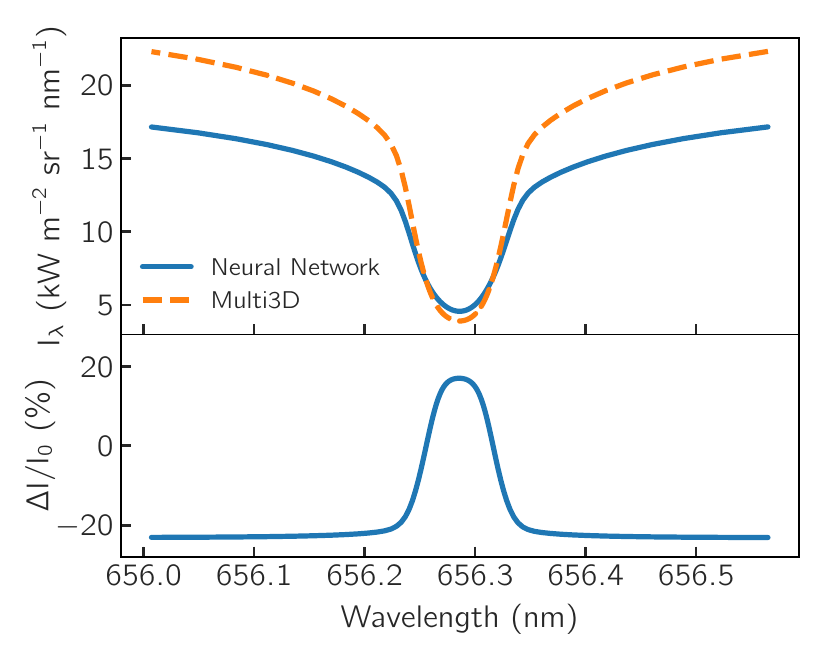}\\
    {\footnotesize (b) High resolution simulation}
    \caption{Spatially averaged H$\alpha$ spectra for the out of sample simulations, covering the wavelength range of $\pm 127\; \mathrm{km}\;\mathrm{s}^{-1}$ around the line core.}
    \label{mixed_diff}
\end{figure}

For the extended simulation, the median normalized difference for all points and levels is 0.699, while the 10th percentile is 0.132 and the 90th percentile 11.583. The $n=6$ level is predicted more accurately than the other levels, whose predictions get worse in the region of average column masses between $10^{-4}-10^{-2}\;\mathrm{kg}\;\mathrm{m}^{-2}$, which is critical for the formation of H$\alpha$. Moreover, the distributions of the normalized differences (not shown) are not as symmetric as for the previous simulations, but instead show a double peaked structure, with the largest peak closer to the median at around 0.5, but with a significant second peak around 5, with a tail extending to more than 100. 

The intensity maps for the extended simulation are shown in Figure~\ref{nw072100}, and the  averaged spectra for both simulations are shown in Figure~\ref{mixed_diff}. As expected from the much worse population estimates, the H$\alpha$ intensities for the SunnyNet predictions show strong departures from the Multi3D results. The line core itself is only slightly underestimated, but is much wider and the wings are strongly underestimated, close to 20\% below the Multi3D values. Both are a reflection of the poor estimates of populations at all regions where the line is formed.

For the second out of sample simulation, the ``high resolution'' simulation, a similar pattern emerges. The estimates from SunnyNet are somewhat worse for the out of sample simulations. For the high resolution simulation, the median normalized difference for all points and levels is 0.726, while the 10th percentile is 0.110 and the 90th percentile 4.404. All levels have poor predictions, with $n=2$ and $n=3$ slightly worse in the H$\alpha$ forming region. The distributions of normalized differences for the different levels are asymmetric and show a double-peaked structure. The predictions show a second peak at higher normalized differences and extending beyond 10. In Figure~\ref{errors_out} we again see that the median normalized differences are much higher than for the ``in sample'' simulations, reaching orders of magnitude off the true result, and particular acute differences at the very top layers and around an average column mass of $10^{-3}\;\mathrm{kg}\;\mathrm{m}^{-2}$. Comparing the accuracy of the predictions against the departure coefficients we find little correlation, as in all the other simulations. The only exception is the $n=1$ level, whose predictions become markedly less accurate for $b>10^8$.

The intensity maps for the high resolution simulation are shown in Figure~\ref{qs006023}. They again illustrate how SunnyNet was not able to accurately predict the populations in regions where H$\alpha$ is formed. Some map aspects are retained, such as granular morphologies, but overall the intensity levels are underestimated by SunnyNet. In the spatially averaged spectra of Figure~\ref{mixed_diff} we see how the line wing intensity is underestimated by more than 20\%. Here, the H$\alpha$ core intensity is overestimated by nearly 20\%.

\section{Discussion}

\subsection{Context}

SunnyNet is a relatively simple convolutional neural network that we used to learn the mapping between LTE and NLTE atomic level populations of a given atomic species for a 3D simulation of a stellar atmosphere. This mapping takes into account how radiative processes affect the populations in 3D. These processes can be strongly nonlocal in the higher atmosphere: when photon mean free paths are larger, the populations at a given cell of the simulation will be affected by the radiation of nearby cells. SunnyNet accounts for these effects by taking as input a window of cells (of a fixed but configurable size) around each cell that we want as output. Once provided with an estimate of the NLTE populations, we can then compute the real goal: synthetic spectra for any line included in the model atom, and for any viewing angle. SunnyNet is therefore one of the first 3D NLTE radiative transfer codes based on machine learning.

Machine learning is today extremely attractive for 3D radiative transfer problems because it can be computationally much cheaper. Radiative transfer in 3D is resource-hungry. Even more for NLTE problems where the angle-averaged radiation field must be evaluated by solving the problem over many rays, and repeatedly for many iterations. With typical problems this needs hundreds or thousands of iterations \citep[see e.g.,][]{Bjorgen:2018} and when adding more complex physics such as partial redistribution it requires hundreds of thousands of CPU-hours for a single simulation snapshot at moderately low spatial resolution \citep{Sukhorukov:2017}. Therefore, the problem is fertile for opportunities to speed it up using machine learning, which can potentially greatly reduce the computational expense. There have been several other ways to approach this problem with machine learning. Earlier work aimed at inverting 1D spectra using NLTE \citep[e.g.,][]{AsensioRamos:2019,osborne}, typically learns the conversion from atmospheric quantities to spectra. Recently, \cite{Vicente:2021} also develop methods for fast synthesis of NLTE spectra, but by learning the conversion from atmospheric parameters to departure coefficients, the ratio between NLTE and LTE populations, which is closer to what we did with SunnyNet. \cite{Mishra:2021} propose an altogether different approach: a machine learning method to learn the actual radiative transfer equation, which has the potential to be much more general and shows very promising results (though it has not been applied to NLTE problems in stellar atmospheres yet).

\subsection{Strengths and limitations}

Running SunnyNet for the six-level hydrogen model atom for a simulation snapshot with $252\times252\times400$ pixels on a single machine with a GPU takes less than one minute, or up to 30 min if we include also the time to train the network and synthesize the H$\alpha$ intensity. This is a speedup of about $10^5$ compared to running Multi3D.

Despite SunnyNet being able to provide estimates of NLTE populations several orders of magnitude faster than conventional methods, it is important to discuss its limitations. The output of SunnyNet is just an estimate of the true solution. Given a prior training, the network tries to find the most likely solution within that base of knowledge that fits the input data. This means that more extreme values will be harder to reproduce. This can be seen in nearly all simulations, but is clearly observed in the H$\alpha$ core maps from the flaring simulation in Figure~\ref{cbh24_0957} -- the brightest regions are not as bright when using the estimate from SunnyNet. More common features such as the dark H$\alpha$ fibrils are reproduced more accurately by SunnyNet. 

When trained and tested in similar simulations, SunnyNet's estimates are reasonably accurate, with most points predicted within 20--40\% of the true result. This may seem large, but the radiative transfer equation is not linearly sensitive to the populations, and the effect on the predicted H$\alpha$ spectra is usually only a few percent. The quality of the predictions is not spatially uniform, not only in terms of more extreme points but also in the smoothness of intensity maps. A close look at the intensity maps, in particular for H$\alpha$ line core shows that the SunnyNet intensities are less smooth than the ones from Multi3D, with a faint but spurious substructure that resembles noise. In any case, this effect is rather subtle. On the other hand, the clear appearance of fibrils in the H$\alpha$ images from SunnyNet is a very encouraging sign, since as \citet{Leenaarts:2012} show, these features appear only when radiative transfer is solved in 3D, and will be totally absent (at least in the enhanced network simulation, which the authors also use) when radiative transfer is performed in 1D. This suggests that the approach followed by SunnyNet is sound, and the 3D effects of radiation are being properly taken into account. 

The tests with out-of-sample simulations show that in its present formulation, SunnyNet performs poorly when testing with different types of simulations that were not used for the training. It is possible this problem could be mitigated by training SunnyNet in a wide range of simulations, therefore providing a much more complete set. But we did not test this, and used only two families of simulations for training before testing with different families. One may also note that the test with the high-resolution simulation was not a fair comparison, since using the same window size would force different physical sizes of the pixel windows and therefore is inconsistent. But numerical simulations come in different resolutions, and it can become time consuming to have complete training sets for all the resolutions used, so it was important to identify how this would affect the results. What is clear is that SunnyNet performs very well at least with the same family of simulations. 

Even in the best-performing cases, SunnyNet is not currently reliable to predict individual spectra with high accuracy. When accuracy is paramount (e.g., element abundances) it cannot replace existing methods. And since it needs to be trained using existing NLTE populations, it makes little sense to use it when analyzing single snapshots of a simulation. However, a possible use case would be to use SunnyNet to compute time series of many snapshots. A few training sets would need to be run with a traditional code such as Multi3D, which could be fed to SunnyNet for computing populations for many snapshots in the same series. SunnyNet can give reasonable estimates of the shapes of spectral lines and very good estimates of intensity maps.

\subsection{Extending SunnyNet}

The experiments so far with SunnyNet show a very positive outcome, but there is nevertheless room for improvement. There are several ways SunnyNet could be extended. One way to do it is to modify the window function to better follow the physics of radiation in 3D. Currently the window consists of using a fixed set of columns around the column of interest, since photons can travel across cells and thereby inclined radiation influences the surrounding regions. However, the photon mean free path varies with height, and in the deeper regions it is typically much smaller than the simulation cell sizes, while near the top it can span many cell sizes. A constant window size is easier to implement numerically, but a more physical scenario would be one where the window size can change with height, like in a cone or extended beam. It could be just one (meaning no effect of nearby cells) in the deep regions, and then increase with height to mimic the increase in photon mean free path. This could fix the inconsistency we found with the window sizes: where too large a window results in worse results. 

Another approach that one could try was to learn not the NLTE populations but the departure coefficients for each level, $b\equiv n_\mathrm{NLTE} / n_\mathrm{LTE}$, similar to what \citet{Vicente:2021} do, but perhaps starting from LTE populations and not the full simulation quantities. This approach could lead to more realistic estimates of the populations at least in deeper regions, which as we saw have similar relative errors than at any other heights, and for physical reasons (matter in LTE when density is high) we would expect them to be easier to estimate. The present version of SunnyNet can be trained with the departure coefficients instead of NLTE populations, since the network is agnostic to the quantity it is fed. We did some preliminary tests running this way, and find that while it gives good estimates, especially improved in the deeper layers, it can also introduce spurious results in several columns. Therefore, it will be necessary to adjust the architecture of the network and tweak the parameters to make it work with the departure coefficients.

Finally, one could also see SunnyNet not as the end goal but as another tool to facilitate the quick calculation of 3D NLTE spectra. The estimates from SunnyNet could be fed to a traditional code such a Multi3D, which would use them as starting guess. Since the SunnyNet populations are much closer to the true result than the LTE populations, they would presumably lead to far fewer iterations needed before achieving convergence. This is a point where we also did some preliminary tests, by feeding the SunnyNet populations into Multi3D. Unfortunately, this did not work directly. Multi3D was unable to converge the SunnyNet populations into a stable solution. A specific reason was not clear, but could be related to the faint spatial substructure seen SunnyNet images. Some more work is needed here, and it could be that one could apply some filtering to the SunnyNet populations to make them more amenable to Multi3D or other codes.

\section{Conclusions}

We demonstrated that a convolutional neural network can be used to accurately predict atomic level populations taking into account the effects of 3D NLTE radiative transfer. These predicted populations can then be used to solve the radiative transfer equation in 3D for any spectral line between the levels predicted, and at any viewing angle. 

Our implementation, SunnyNet, uses the PyTorch library, which can take advantage of GPUs to greatly speed up calculations. Compared to traditional codes to solve 3D NLTE radiative transfer, SunnyNet is about $10^5$ times faster. These gains in running time come at the expense of some accuracy, so the results from SunnyNet should not be seen as the ground truth but instead as a fast approximation.

Using a model atom of hydrogen and synthesizing the H$\alpha$ line, we showed that SunnyNet's predicted populations are for most points within 20--40\% of the true result, which for the H$\alpha$ line translates to mean differences of only a few percent in the line core, and smaller differences in the line wings. The overall morphology of the intensity maps (synthetic spectroheliograms) computed from SunnyNet's predictions shows very good agreement with those from Multi3D. In particular, the telltale signs of 3D radiative transfer are present in the maps from SunnyNet. The appearance of fibrils and loops in H$\alpha$, which do not appear in 1.5D radiative transfer, suggests that SunnyNet is correctly learning the 3D effects of NLTE radiative transfer. SunnyNet works equally well in regions with strong departures from LTE, and for various Bifrost simulations, ranging from quiet Sun to more dynamic and showing small flares.

SunnyNet works best when training the network with snapshots of the same simulation that we want to compute the spectra from. Training with a couple of snapshots from a given simulation gives good results to predict the spectra of other snapshots from the same simulation. Using SunnyNet to predict populations from a simulation different from the one used for the training does not give reliable results. It is possible this could be mitigated by building a larger training database comprising multiple snapshots from a variety of simulations, but we did not test this (our tests extended only to a total of four snapshots from two different simulations). 

SunnyNet is not yet able to completely replace existing methods for 3D NLTE radiative transfer. Traditional codes are still necessary to compute the populations needed to train SunnyNet, and when accuracy is paramount. SunnyNet could be used to provide fast computations of time series, which are prohibitively expensive using current methods, and then a traditional code could be used to further refine the results in regions of interest. Another alternative is to use SunnyNet, well trained with different simulations, to instead provide the starting guess of populations to traditional codes, which could then converge faster to the final populations. There are several points where SunnyNet can be improved or extended, and we believe similar techniques will be in great demand in the future.

\begin{acknowledgements}
The authors thank V. H. Hansteen and M. Carlsson for providing Bifrost simulations and for fruitful discussions. This work has been supported by the Research Council of Norway through its Centres of Excellence scheme, project number 262622. Computational resources have been provided by UNINETT Sigma2 - the National Infrastructure for High Performance Computing and Data Storage in Norway and by the High End Computing (HEC) division of NASA, Pleiades cluster, through computing projects s1061 and s8305. Running the extended simulation was made possible by support from NASA grant 19-HTMS19 2-0025 ``Flux emergence and the structure, dynamics, and energetics of the solar atmosphere''.
\end{acknowledgements}

\bibliographystyle{aa} %
\bibliography{references} %

\end{document}